\documentclass[english,preprint]{aastex}
\usepackage[T1]{fontenc}
\usepackage[latin9]{inputenc}
\usepackage{amsmath}
\usepackage{amssymb}

\makeatletter



\makeatother

\usepackage{babel}

\begin{document}

\title{Dispersion of Magnetic Fields in Molecular Clouds. III}

\author{Martin Houde$^{1,2}$, Ramprasad Rao$^{3}$, John E. Vaillancourt$^{4}$,
and Roger H. Hildebrand$^{5,6}$}

\affil{$^{1}$Department of Physics and Astronomy, The University of Western
Ontario, London, ON, N6A 3K7, Canada}

\affil{$^{2}$Division of Physics, Mathematics and Astronomy, California
Institute of Technology, Pasadena, CA 91125 }

\affil{$^{3}$Academia Sinica Institute of Astronomy and Astrophysics, Taipei,
Taiwan}

\affil{$^{4}$Stratospheric Observatory for Infrared Astronomy, Universities
Space Research Association, NASA Ames Research Center, Moffet Field,
CA 94035}

\affil{$^{5}$Department of Astronomy and Astrophysics and Enrico Fermi
Institute, The University of Chicago, Chicago, IL 60637}

\affil{$^{6}$Department of Physics, The University of Chicago, Chicago,
IL 60637}
\begin{abstract}
We apply our technique on the dispersion of magnetic fields in molecular
clouds to high spatial resolution Submillimeter Array polarization
data obtained for Orion KL in OMC-1, IRAS 16293, and NGC 1333 IRAS
4A. We show how one can take advantage of such high resolution data
to characterize the magnetized turbulence power spectrum in the inertial
and dissipation ranges. For Orion KL we determine that in the inertial
range the spectrum can be approximately fitted with a power law $k^{-\left(2.9\pm0.9\right)}$
and we report a value of 9.9 mpc for $\lambda_{\mathrm{AD}}$, the
high spatial frequency cutoff presumably due to turbulent ambipolar
diffusion. For the same parameters we have $\sim k^{-\left(1.4\pm0.4\right)}$
and a tentative value of $\lambda_{\mathrm{AD}}\simeq2.2$ mpc for
NGC 1333 IRAS 4A, and $\sim k^{-\left(1.8\pm0.3\right)}$ with an
upper limit of $\lambda_{\mathrm{AD}}\lesssim1.8$ mpc for IRAS 16293.
We also discuss the application of the technique to interferometry
measurements and the effects of the inherent spatial filtering process
on the interpretation of the results. 
\end{abstract}


\section{Introduction}

The fact that the relative importance of magnetic fields and turbulence
in the process of star formation is still a matter of debate \citep{Mouschovias2009,Crutcher2010}
can be traced to the many difficulties of probing magnetic fields
in molecular clouds. While the Zeeman effect still provides the only
direct way of measuring the strength of (generally the line of sight
component of) magnetic fields \citep{Heiles1997,Crutcher1999,Brogan2001,Falgarone2008},
its weakness in the interstellar medium limits the types of environments
and number of regions where detections can successfully be obtained.
Without the existence of a prolific technique for measuring the strength
of all components of the magnetic field vector, it is difficult to
precisely quantify its importance. It is therefore imperative to keep
seeking new ideas and techniques that could provide information, even
partially, on the nature of magnetic fields. Interestingly, such newly
introduced methods of observation or analyses that have been recently
put forth for the study of the magnetic field do so by taking advantage
of its interplay with turbulence \citep{Houde2000,Li2008,Heyer2008}. 

In this paper we continue our previous work on the characterization
of magnetized turbulence in molecular clouds using polarization maps.
In \citet{Hildebrand2009} (hereafter Paper I) we showed how the turbulent
to ordered magnetic field strength ratio can be evaluated through
the structure function of the polarization angles (i.e., the dispersion
function) without assuming any model for the ordered component of
the magnetic field. Subsequently in \citet{Houde2009} (hereafter
Paper II) we generalized this analysis by including the process of
signal integration through the thickness of the cloud and across the
telescope beam. An important development of Paper II was the determination
of the magnetized turbulent correlation length scale, which is in
effect a measure for the width of the magnetized turbulence power
spectrum. We go further in this paper by taking advantage of high
resolution polarization maps to characterize the magnetized turbulence
power spectrum in the inertial and dissipation ranges.

We will start in Section \ref{sec:Analysis} with a brief review of
the material presented in Paper II on the cloud- and beam-integrated
dispersion function that is necessary for our analysis. In Section
\ref{sec:Observations} we present the data on which we will perform
our analysis: high resolution SMA polarization data of Orion KL (similar
to those used for the map presented in Fig. 2 c) of \citealt{Tang2010}),
as well as the previously published SMA data of NGC 1333 IRAS 4A \citep{Girart2006}
and IRAS 16293 \citep{Rao2009}. We present the results of our dispersion
function analysis in Section \ref{sec:Results}, where we emphasize
some important considerations that are specific to interferometry
data. We then follow up with a discussion on the limitations of our
technique, improvements that could be implemented, and future applications
in Section \ref{sec:Discussion}. We end with a short summary in Section
\ref{sec:Summary}. Finally, an appendix pertaining to some data processing
aspects can be found at the end of the paper.

\section{The Magnetized Turbulence Power Spectrum \label{sec:Analysis}}

Following the analysis of Paper I and Paper II we model the dispersion
of the difference $\Delta\Phi\left(\boldsymbol{\ell}\right)\equiv\Phi\left(\mathbf{x}\right)-\Phi\left(\mathbf{x}+\boldsymbol{\ell}\right)$
in the polarization angle $\Phi$ measured at two positions separated
by a distance $\boldsymbol{\ell}$ with

\begin{equation}
\left\langle \cos\left[\Delta\Phi\left(\ell\right)\right]\right\rangle =\frac{\left\langle \overline{\mathbf{B}}\mathbf{\cdot}\overline{\mathbf{B}}\mathbf{\left(\ell\right)}\right\rangle }{\left\langle \overline{\mathbf{B}}\mathbf{\cdot}\overline{\mathbf{B}}\left(0\right)\right\rangle },\label{eq:cos}\end{equation}

\noindent where $\left\langle \cdots\right\rangle $ denotes an average
and $\ell=\left|\boldsymbol{\ell}\right|$ and $\left\langle \overline{\mathbf{B}}\mathbf{\cdot}\overline{\mathbf{B}}\mathbf{\left(\ell\right)}\right\rangle \equiv\left\langle \overline{\mathbf{B}}\left(\mathbf{r}\right)\mathbf{\cdot}\overline{\mathbf{B}}\mathbf{\left(\mathbf{r}+\boldsymbol{\ell}\right)}\right\rangle $
(see below). The cloud- and beam-integrated magnetic field is defined
with

\begin{equation}
\overline{\mathbf{B}}\left(\mathbf{r}\right)=\iint H\left(\mathbf{r}-\mathbf{a}\right)\left[\frac{1}{\Delta}\int_{0}^{\Delta}F\left(\mathbf{a},z\right)\mathbf{B}\left(\mathbf{a},z\right)dz\right]d^{2}a,\label{eq:Bbar}\end{equation}

\noindent where the beam profile is denoted by $H\left(\mathbf{r}\right)$,
while the weighting function $F\left(\mathbf{a},z\right)\geq0$ is
the polarized emission associated with the magnetic field $\mathbf{B}\left(\mathbf{a},z\right)$
and $\Delta$ is the maximum depth of the cloud along any line of
sight. The quantity $\mathbf{r}$ is the two-dimensional polar radius
vector on the plane-of-the-sky and $z$ the depth within the cloud.
That is, the position vector in the cloud is given by

\begin{equation}
\mathbf{x}=r\mathbf{e}_{r}+z\mathbf{e}_{z}\label{eq:x}\end{equation}

\noindent with $\mathbf{e}_{r}$ and $\mathbf{e}_{z}$ the unit basis
vectors along $\mathbf{r}$ and the $z$-axis (which is oriented along
the line of sight), respectively. The distance $\ell$ in Equation
(\ref{eq:cos}) is also confined to the plane of the sky. We assume
that the magnetic field $\mathbf{B\left(\mathbf{x}\right)}$ is composed
of an ordered field, $\mathbf{B}_{0}\mathbf{\left(\mathbf{x}\right)}$,
and a turbulent (random), zero-mean component, $\mathbf{B_{\mathrm{t}}\left(\mathbf{x}\right)}$,
such that 

\begin{equation}
\mathbf{\mathbf{B\left(\mathbf{x}\right)}}=\mathbf{B}_{0}\mathbf{\mathbf{\mathbf{\left(\mathbf{x}\right)+\mathbf{B}_{\mathrm{t}}\left(\mathbf{x}\right)}}}.\label{eq:Btot}\end{equation}

\noindent For simplicity, we assume the polarized flux $F\left(\mathbf{x}\right)$
to be composed of an ordered component only (the more general case
where a turbulent component is added was discussed in Paper II). Stationarity,
homogeneity and isotropy (see Sec. \ref{sub:future}) in the magnetic
field strength were assumed for Equation (\ref{eq:cos}), while statistical
independence between ordered and turbulent components will also be
implied in what follows.

Following the analysis of Paper II it is found that the dispersion
function $1-\left\langle \cos\left[\Delta\Phi\left(\ell\right)\right]\right\rangle $
can be broken up into turbulent and ordered terms 

\begin{eqnarray}
1-\left\langle \cos\left[\Delta\Phi\left(\ell\right)\right]\right\rangle  & = & \left[b^{2}\left(0\right)-b^{2}\left(\ell\right)\right]+\left[\alpha^{2}\left(0\right)-\alpha^{2}\left(\ell\right)\right]\nonumber \\
 & = & \left\{ b^{2}\left(0\right)+\left[\alpha^{2}\left(0\right)-\alpha^{2}\left(\ell\right)\right]\right\} -b^{2}\left(\ell\right),\label{eq:b_alpha}\end{eqnarray}

\noindent with the ordered and turbulent autocorrelation functions
given by 

\begin{eqnarray}
\alpha^{2}\left(\ell\right) & = & \frac{\left\langle \overline{\mathbf{B}}_{0}\mathbf{\cdot}\overline{\mathbf{B}}_{0}\left(\ell\right)\right\rangle }{\left\langle \overline{\mathbf{B}}\mathbf{\cdot}\overline{\mathbf{B}}\left(0\right)\right\rangle }\label{eq:alpha}\\
b^{2}\left(\ell\right) & = & \frac{\left\langle \overline{\mathbf{B}}_{\mathrm{t}}\mathbf{\cdot}\overline{\mathbf{B}}_{\mathrm{t}}\left(\ell\right)\right\rangle }{\left\langle \overline{\mathbf{B}}\mathbf{\cdot}\overline{\mathbf{B}}\left(0\right)\right\rangle },\label{eq:b2}\end{eqnarray}

\noindent respectively. The ordered function $\alpha^{2}\left(0\right)-\alpha^{2}\left(\ell\right)$
can be advantageously modeled with a Taylor series 

\begin{equation}
\alpha^{2}\left(0\right)-\alpha^{2}\left(\ell\right)=\sum_{j=1}^{\infty}a_{2j}\ell^{2j},\label{eq:Taylor}\end{equation}

\noindent while $b^{2}\left(0\right)$ is simply the turbulent to
total magnetic energy ratio (that is, for the corresponding cloud-
and beam-integrated quantities). As was shown in Paper I and Paper
II the quantity within curly braces in Equation (\ref{eq:b_alpha})
can be readily determined from polarization maps by calculating the
dispersion function (i.e., the left-hand side of that equation) from
the data and fitting

\begin{equation}
b^{2}\left(0\right)+\left[\alpha^{2}\left(0\right)-\alpha^{2}\left(\ell\right)\right]=b^{2}\left(0\right)+\sum_{j=1}^{\infty}a_{2j}\ell^{2j}\label{eq:model}\end{equation}

\noindent to the data outside of the region where $b^{2}\left(\ell\right)$
is dominant (i.e., at lower values of $\ell$). Once Equation (\ref{eq:model})
is evaluated the (normalized) turbulent cloud- and beam-integrated
autocorrelation function $b^{2}\left(\ell\right)$ can be extracted
from the data through Equation (\ref{eq:b_alpha}). 

Alternatively, the integrated turbulent autocorrelation function $b^{2}\left(\ell\right)$
can also be analytically derived using (see Eq. {[}A5{]} of Paper
II)

\begin{equation}
\left\langle \overline{\mathbf{B}}_{\mathrm{t}}\mathbf{\cdot}\overline{\mathbf{B}}_{\mathrm{t}}\mathbf{\left(\boldsymbol{\ell}\right)}\right\rangle =\iint\iint H\left(\mathbf{a}\right)H\left(\mathbf{a}^{\prime}+\boldsymbol{\ell}\right)\left[\frac{2}{\Delta}\int_{0}^{\Delta}\left(1-\frac{u}{\Delta}\right)\mathcal{R}_{\mathrm{3\mathrm{D},t}}\left(v,u\right)du\right]d^{2}a^{\prime}d^{2}a,\label{eq:autob2}\end{equation}

\noindent with $\mathcal{R}_{\mathrm{3\mathrm{D},t}}\left(v,u\right)=\left\langle F\left(\mathbf{a},z\right)F\left(\mathbf{a}^{\prime},z^{\prime}\right)\right\rangle \left\langle \mathbf{B}_{\mathrm{t}}\left(\mathbf{a},z\right)\cdot\mathbf{B}_{\mathrm{t}}\left(\mathbf{a}^{\prime},z^{\prime}\right)\right\rangle $,
$u=\left|z^{\prime}-z\right|$ and $v=\left|\mathbf{a}^{\prime}-\mathbf{a}\right|$,
and a similar equation for $\left\langle \overline{\mathbf{B}}\mathbf{\cdot}\overline{\mathbf{B}}\left(0\right)\right\rangle $.
Since we are mostly interested in determining the shape of the magnetized
turbulence power spectrum, we will concentrate on the Fourier transform
of $b^{2}\left(\ell\right)$ (see Eq. {[}A12{]} of Paper II)

\begin{equation}
b^{2}\left(\mathbf{k}_{v}\right)=\frac{1}{\left\langle \overline{B}^{2}\right\rangle }\left\Vert H\left(\mathbf{k}_{v}\right)\right\Vert ^{2}\left[\int\mathcal{R}_{\mathrm{3\mathrm{D},t}}\left(\mathbf{k}_{v},k_{u}\right)\mathrm{sinc}^{2}\left(\frac{k_{u}\Delta}{2}\right)dk_{u}\right],\label{eq:b2(kv)_int}\end{equation}

\noindent where $\left\langle \overline{B}^{2}\right\rangle \equiv\left\langle \overline{\mathbf{B}}\mathbf{\cdot}\overline{\mathbf{B}}\left(0\right)\right\rangle $
and the Fourier transform of a function is represented by simply replacing
the spatial arguments by their $\mathbf{k}$-space counterparts (e.g.,
$\mathcal{R}_{\mathrm{3\mathrm{D},t}}\left(v,u\right)\rightleftharpoons\mathcal{R}_{\mathrm{3\mathrm{D},t}}\left(\mathbf{k}_{v},k_{u}\right)$).
Equation (\ref{eq:b2(kv)_int}) can be redefined with

\begin{equation}
b^{2}\left(\mathbf{k}_{v}\right)=\left\Vert H\left(\mathbf{k}_{v}\right)\right\Vert ^{2}\frac{\mathcal{R}_{\mathrm{t}}\left(\mathbf{k}_{v}\right)}{\left\langle \overline{B}^{2}\right\rangle },\label{eq:b2(kv)}\end{equation}

\noindent where $\mathcal{R}_{\mathrm{t}}\left(\mathbf{k}_{v}\right)\equiv\int\mathcal{R}_{\mathrm{3\mathrm{D},t}}\left(\mathbf{k}_{v},k_{u}\right)\mathrm{sinc}^{2}\left(k_{u}\Delta/2\right)dk_{u}$
is now interpreted as the two-dimensional turbulence power spectrum
we seek to evaluate. We will accomplish this by taking the Fourier
transform of $b^{2}\left(\ell\right)$ obtained from the data, as
explained in the discussion following Equation (\ref{eq:model}) above,
to evaluate $b^{2}\left(\mathbf{k}_{v}\right)$ on the left-hand side
of Equation (\ref{eq:b2(kv)}) and then invert this relation to determine
$\mathcal{R}_{\mathrm{t}}\left(\mathbf{k}_{v}\right)/\left\langle \overline{B}^{2}\right\rangle $
(with a Wiener filter to remove the filtering due to $\left\Vert H\left(\mathbf{k}_{v}\right)\right\Vert ^{2}$;
see Appendix).

\section{Observations\label{sec:Observations}}

The observations for Orion KL were carried out on 10 September 2006
and 6 January 2008 using the SMA \citep{Ho2004}%
\footnote{The Submillimeter Array is a joint project between the Smithsonian
Astrophysical Observatory and the Academia Sinica Institute of Astronomy
and Astrophysics and is funded by the Smithsonian Institution and
the Academia Sinica.%
} in the compact array configuration, with the projected baseline lengths
ranging from 15 to 80 k$\lambda$ ($\lambda=870\,\mu$m). The phase
center is at $\mathrm{RA}(\mathrm{J2000})=5^{\mathrm{h}}35^{\mathrm{m}}14\fs5$,
$\mathrm{Decl}(\mathrm{J2000})=-5^{\circ}22\arcmin30\farcs4$. The
SMA receiver system has two sidebands, each with a bandwidth of $\sim2$
GHz. The sampled sky frequencies range from 345.5 to 347.5 GHz in
the upper sideband and from 335.5 to 337.5 GHz in the lower, with
a uniform spectral resolution of 0.812 MHz (corresponding to a velocity
resolution of 0.7 km s$^{-1}$). At these frequencies, the primary
beam size (or field of view) of the SMA is $\sim32\arcsec$. Within
the observational bandwidth, there is a significant contribution to
the total emission from spectral lines of a number of molecular transitions
(notably CO $\left(J=3\rightarrow2\right)$ and SiO $\left(J=8\rightarrow7\right)$),
and the continuum is generated after removing the spectral line contamination.
Using natural weighting of the visibilities, the synthesized beam
size is $2\farcs6\times1\farcs7$. The noise level in the Stokes $I$
image is $\sim0.3$ Jy beam$^{-1}$. This is much higher than the
theoretical noise level due to the limited dynamic range in the Stokes
$I$ map. The noise levels of both the Stokes $Q$ and $U$ images,
which are much fainter, are much closer to the theoretical noise level,
at 10 mJy beam$^{-1}$. Our observations for Orion KL have much data
in common with those used for the polarization map presented in Figure
2 c) of \citet{Tang2010}, which we refer the reader to in view of
its similarity to the map that can be derived from our data. We note,
however, that our data have a slightly higher spatial resolution than
the $2\farcs8\times1\farcs8$ synthesized beam size of \citet{Tang2010}.
\citet{Koch2010} also performed a dispersion analysis on the \citet{Tang2010}
map based on our Papers I and II. Importantly for our analysis, we
use all polarization vectors available at the sampling rate of $0\farcs25$,
provided they satisfy the condition $p\geq3\sigma_{p}$, with $p$
and $\sigma_{p}$ the polarization level and its uncertainty, respectively.
This implies that many of our data points are correlated with each
other since several are contained with a single synthesized beam profile.
This is beneficial for our analysis as it allows for a better fit
of Equation (\ref{eq:model}) from the dispersion data. The correlation
between data points is also accounted for in the evaluation of the
dispersion function and its uncertainty as a function of $\ell$ (see
Appendix B of Paper II).

The data were calibrated and processed using the software package
MIRIAD \citep{Wright1993}. The gain calibration was obtained from
observations of the QSO 0528+134. It is necessary to remove the contributions
due to instrumental polarization as these are roughly similar in magnitude
to the observed source polarization and can corrupt the data (see
\citealt{Marrone2006} and \citealt{Marrone2008} for the details
of this method). The instrumental polarization was obtained from observations
of the strong quasar 3c273 for 2 hours during transit. The total intensity
(Stokes $I$) map was deconvolved using the task CLEAN in MIRIAD.
We derived the polarized intensity ($I_{\mathrm{p}}$) and position
angles (P.A.s) with the task IMPOL, also using the CLEANed Stokes
$Q$ and $U$ maps. The task IMPOL further removed the effects of
the bias of the positive measure of $I_{\mathrm{p}}$. 

The data for IRAS 16293 and NGC 1333 IRAS 4A were also obtained at
the SMA and reduced in a similar manner. They were previously published
and described in detail in \citet{Rao2009} and \citet{Girart2006},
respectively. Their synthesized beam sizes and sampling rates are
$3\farcs1\times2\farcs0$ and $0\farcs25$ for IRAS 16293, and $1\farcs6\times1\farcs0$
and $0\farcs2$ for NGC 1333 IRAS 4A.

\section{Results\label{sec:Results}}

\subsection{Dispersion Functions from Interferometry Data - Orion KL}

Although the discussion that follows applies equally well to all the
data presented in this paper, we will refer only to the Orion KL data
(see map of Figure 2 c) of \citealt{Tang2010}) in this sub-section
in order to discuss the application of the dispersion analysis to
interferometry data. One of our main goals here is to emphasize that
there are important differences with results obtained with interferometry
in relation to what one would get with single-dish data. 

We first note that because of the unavoidable 180-degree ambiguity
when determining the orientation of the magnetic field from polarization
data, one has to be careful when proceeding with a dispersion analysis
of polarization angles. Indeed, any such analysis can only be successfully
carried out in regions of polarization maps for which changes in polarization
angles with position are sufficiently smooth to ensure, at least to
a reasonable degree, that there are no possible reversals in the field
direction. For this reason, we have excluded from our analysis a small
isolated {}``clump'' located at an offset of $\Delta\alpha\approx8\arcsec$
and $\Delta\delta\approx11\arcsec$ to the north of position of peak
intensity of Orion KL (i.e., at $\alpha\left(\mathrm{J2000}\right)\simeq5^{\mathrm{h}}35^{\mathrm{m}}14\fs9$
and $\delta\left(\mathrm{J2000}\right)\simeq-5^{\circ}22\arcmin17\arcsec$
in Figure 2 c) of \citealt{Tang2010}). 

Following the analysis done in Paper II, the dispersion function $1-\left\langle \cos\left[\Delta\Phi\left(\ell\right)\right]\right\rangle $
and turbulent autocorrelation function $b^{2}\left(\ell\right)$ determined
from the Orion KL polarization map are shown as a function of the
distance $\ell$ in Figure \ref{fig:omc-1_struct} (symbols in top
and bottom graphs, respectively). It is important to note that although
these functions are only plotted for $\ell\geq0$, the data is actually
two-dimensional in nature and exhibits cylindrical symmetry (in the
plane containing $\ell$ about the $\ell=0$ axis) because of the
assumed isotropy of the dispersion function. With the prescription
given in Paper I (or Paper II) one then fits the (sum of the) turbulent
to total magnetic energy ratio $b^{2}\left(0\right)$ and the ordered
magnetic field component ${\textstyle \sum_{j}}a_{2j}\ell^{2j}$,
for which we use the lowest order polynomial that fits the data, with
the broken curve shown in the top graph (see Equation {[}\ref{eq:model}{]});
the subtraction of the data from that curve would then yield $b^{2}\left(\ell\right)$
from which the analysis could proceed (see Equation {[}\ref{eq:b_alpha}{]}).
Although this is perfectly adequate for single-dish data, an important
fact needs to be emphasized when dealing with interferometry data. 

The aforementioned subtraction of the dispersion data from the broken
curve in the top graph of Figure \ref{fig:omc-1_struct} leads to
turbulent autocorrelation data that satisfy $b^{2}\left(\ell\right)\geq0$
for most if not all values of $\ell$, which also implies, of course,
that the integral of that function over space will be positive. But
this should not be possible with a polarization map obtained with
an interferometer, as can be assessed from Equation (\ref{eq:b2(kv)}),
since 

\begin{eqnarray}
2\pi\int_{0}^{\infty}b^{2}\left(\ell\right)\ell d\ell & = & b^{2}\left(\mathbf{k}_{v}=0\right)\nonumber \\
 & = & \frac{1}{\left\langle \overline{B}^{2}\right\rangle }\left[\left\Vert H\left(\mathbf{k}_{v}\right)\right\Vert ^{2}\mathcal{R}_{\mathrm{t}}\left(\mathbf{k}_{v}\right)\right]_{\mathbf{k}_{v}=0}\label{eq:area}\\
 & = & 0\end{eqnarray}

\noindent because $H\left(\mathbf{k}_{v}=0\right)=0$ for the so-called
{}``dirty beam'' of an interferometer. This, of course, is directly
related to the well-known missing-flux {}``issue'' that is implicit
to interferometry data. But importantly, it is also contrary to single-dish
data where $H\left(\mathbf{k}_{v}=0\right)=1$, when inefficiencies
are accounted for. It therefore follows that an acceptable fit $b^{2}\left(0\right)+{\textstyle \sum_{j}}a_{2j}\ell^{2j}$
to our dispersion function interferometry data should satisfy the
condition

\begin{equation}
2\pi\int_{0}^{\infty}b^{2}\left(\ell\right)\ell d\ell=0.\label{eq:condition}\end{equation}

Although we have restricted the condition of Equation (\ref{eq:condition})
to the turbulent component $b^{2}\left(\ell\right)$, it also applies
equally well to the ordered component and the total normalized autocorrelation
function of the magnetic field given by Equation (\ref{eq:cos}).
The important point we need to acknowledge is that it is impossible
to know exactly what correct fit for the turbulent to total magnetic
energy ratio and ordered component (i.e., $b^{2}\left(0\right)+{\textstyle \sum_{j}}a_{2j}\ell^{2j}$)
applies to our Orion KL data (or any other data set) that will verify
Equation (\ref{eq:condition}), as there is an infinite number of
combinations for $b^{2}\left(0\right)$ and ${\textstyle \sum_{j}}a_{2j}\ell^{2j}$
that will satisfy this condition and no way to discriminate between
them. We must therefore accept this as a fundamental limitation to
the analysis when dealing with interferometry data alone. This problem
can only be avoided if the polarization map includes data that fill
the low-frequency portion of the spectrum (including $\mathbf{k}_{v}=0$),
i.e., when single-dish data are available, since the condition given
in Equation (\ref{eq:condition}) will then not apply anymore and
the degeneracy on the aforementioned fit will be lifted.

On the other hand, it is perhaps reasonable to expect that our fit
for $b^{2}\left(0\right)+{\textstyle \sum_{j}}a_{2j}\ell^{2j}$ (as
shown in Fig. \ref{fig:omc-1_struct}, for example) will affect the
shape of the turbulence power spectrum $\mathcal{R}_{\mathrm{t}}\left(\mathbf{k}_{v}\right)/\left\langle \overline{B}^{2}\right\rangle $,
which we seek to determine from the dispersion function, mainly at
low spatial frequencies. More precisely, the spectral content associated
with the fit $b^{2}\left(0\right)+{\textstyle \sum_{j}}a_{2j}\ell^{2j}$
is to a large extent concentrated at low frequencies and will have
a diminishing effect on our determination of the shape of the magnetized
turbulence power spectrum when moving to higher frequencies. For the
purpose of the present analysis, we will proceed by neglecting the
condition stated in Equation (\ref{eq:condition}) and perform our
study as we normally would for single-dish data, and then discard
the low frequency portion of the turbulent spectrum from our analysis
(see below). That is, we adopt the fit shown in the top graph of Figure
\ref{fig:omc-1_struct} (i.e., the broken curve) for Orion KL for
our analysis. Accordingly, the associated turbulent autocorrelation
function $b^{2}\left(\ell\right)$ is that shown in the bottom graph
of Figure \ref{fig:omc-1_struct} (symbols). Also plotted is the radial
profile of the {}``mean autocorrelated beam'' (broken curve). This
profile is obtained by first computing the autocorrelation of the
synthesized beam, since it is this beam that is used for the polarization
map and appears in the expression for the dispersion function (see
Eqs. {[}\ref{eq:cos}{]} and {[}\ref{eq:autob2}{]}), and then averaged
azimuthally in the same manner as are the dispersion data. This represents
the contribution of the synthesized beam to the (width of) the turbulent
autocorrelation function $b^{2}\left(\ell\right)$. That is, this
is what $b^{2}\left(\ell\right)$ would look like in the limit where
the intrinsic turbulent correlation length were zero. A comparison
of this autocorrelated beam profile to that of $b^{2}\left(\ell\right)$
clearly shows the significant contribution of the magnetized turbulence
$\mathcal{R}_{\mathrm{t}}\left(\ell\right)$ to the overall width
and shape of $b^{2}\left(\ell\right)$.

The top graph of Figure \ref{fig:omc-1_spectrum} shows the spectra
associated with $b^{2}\left(\ell\right)$ (i.e., $b^{2}\left(k\right)$
with $k=\left|\mathbf{k}_{v}\right|$; symbols) and the mean autocorrelated
synthesized beam (i.e., $\left\Vert H\left(k\right)\right\Vert ^{2}$;
broken curve) calculated by taking the Fourier transform of the corresponding
functions shown in the bottom graph of Figure \ref{fig:omc-1_struct}
(see Appendix). Also shown is the spectrum of the visibility data
(or dirty beam; dot-broken curve), which is normalized to its peak
level (using the scale on the right). We can now readily verify that
the telescope dirty beam is responsible for the condition stated through
Equation (\ref{eq:condition}) since it vanishes at $k=0$, as expected.
The fact that our data for $b^{2}\left(k\right)$ are non-zero at
and close to $k=0$ is a reflection of the fact that, as we discussed
above, we cannot precisely evaluate it from the dispersion function.
Because of this we exclude the first two points of the spectrum at
$k/2\pi=0$ and $0.5$ arcsec$^{-1}$ from our analysis and concentrate
on the rest of the spectrum (i.e., higher frequencies). 

The bottom graph of Figure \ref{fig:omc-1_spectrum} presents the
results of our analysis for Orion KL. First, Equation (\ref{eq:b2(kv)})
is inverted using a simple Wiener optimal filter (see Appendix), which
is possible since the beam profile $\left\Vert H\left(k\right)\right\Vert ^{2}$
is well characterized, to yield the magnetized turbulence power spectrum
profile $\mathcal{R}_{\mathrm{t}}\left(k\right)/\left\langle \overline{B}^{2}\right\rangle $
(again with $k=\left|\mathbf{k}_{v}\right|$; symbols). 

The turbulence power spectrum is not usually expressed with $\mathcal{R}_{\mathrm{t}}\left(k\right)$,
however, but rather with \citep{Frisch1995}

\begin{equation}
\mathcal{R}_{\mathrm{K}}\left(k^{\prime}\right)\equiv4\pi k^{\prime2}\mathcal{R}_{\mathrm{3\mathrm{D},t}}\left(k^{\prime}\right),\label{eq:R_K}\end{equation}

\noindent which is a {}``one-dimensional'' representation of (the
three-dimensional) $\mathcal{R}_{\mathrm{3\mathrm{D},t}}\left(\mathbf{k}_{v},k_{u}\right)$
with $k^{\prime2}=\left|\mathbf{k}_{v}\right|^{2}+k_{u}^{2}$. This
{}``Kolmo\-go\-rov-like'' spectrum is particularly well-suited
for cases of isotropic turbulence. It is unfortunately not possible
with our data to recover $\mathcal{R}_{\mathrm{K}}\left(k^{\prime}\right)$
from $\mathcal{R}_{\mathrm{t}}\left(k\right)$ because of the integration
over $k_{u}$ in Equation (\ref{eq:b2(kv)_int}) that is inherent
to the measurement process. On the other hand, it is still possible
to define another {}``one-dimensional'' power spectrum $\mathcal{R}_{\mathrm{1D}}\left(k\right)$
with

\begin{equation}
\mathcal{R}_{1\mathrm{D}}\left(k\right)\equiv2\pi k\mathcal{R}_{\mathrm{t}}\left(k\right),\label{eq:R_1D(k)}\end{equation}

\noindent which is obtained from our data in a straightforward manner.
We show the corresponding result for Orion KL with the solid curve
in the bottom graph of Figure \ref{fig:omc-1_spectrum}. Since it
is also customary to parametrize the turbulence power spectrum in
the inertial range with a power law, we also plot such a fit to $\mathcal{R}_{\mathrm{1D}}\left(k\right)/\left\langle \overline{B}^{2}\right\rangle $
and show that the inertial range approximately scales with $k^{-\left(2.9\pm0.9\right)}$.
Although this result is consistent with theoretical expectations,
the small number of spectral points available for the fit and our
lack of knowledge concerning the precise shape of the ordered component
to be subtracted from the dispersion function reduces the robustness
of our determination for the spectral index (as is exemplified by
the significant uncertainty on its value). 

Another important parameter that characterizes the magnetized turbulence
power spectrum is the cutoff frequency $k_{\mathrm{AD}}$, or length
scale $\lambda_{\mathrm{AD}}=2\pi/k_{\mathrm{AD}}$, at high frequencies,
which is likely due to ambipolar diffusion \citep{Li2008, Hezareh2010, Lazarian2004, Diego2010, Tilley2010, Houde2011}.
We find from our results for $\mathcal{R}_{1\mathrm{D}}\left(k\right)/\left\langle \overline{B}^{2}\right\rangle $
in Figure \ref{fig:omc-1_spectrum} that $k_{\mathrm{AD}}/2\pi\sim0.2$
arcsec$^{-1}$. Although this value ($\lambda_{\mathrm{AD}}\sim11$
mpc, see below) is also consistent with theoretical expectations \citep{Lazarian2004, Diego2010, Tilley2010}
as well as observationally determined values \citep{Li2008, Hezareh2010},
we should ensure that this spectral cut off is real and not artificially
imposed by data processing or the finite spatial resolution of the
interferometer. This can be verified with Figure \ref{fig:omc-1_ad}
where we have magnified the vertical scale of the top graph of Figure
\ref{fig:omc-1_spectrum}. It is observed that this spectral cut-off
is present in the data for $b^{2}\left(k\right)$ (i.e., before applying
the Wiener filter) and seen to happen well within the bandwidth subtended
by the synthesized beam (or rather that of $\left\Vert H\left(k\right)\right\Vert ^{2}$),
which cuts off at $k/2\pi\simeq0.6$ arcsec$^{-1}$. This plot also
shows that $k_{\mathrm{AD}}/2\pi\simeq0.22$ arcsec$^{-1}$, or alternatively
$\lambda_{\mathrm{AD}}\simeq9.9$ mpc for Orion KL (assumed to be
at a distance of 450 pc).

\subsection{IRAS 16293}

We applied our dispersion analysis to the polarization map of IRAS
16293 of \citet{Rao2009}. The results for the dispersion and turbulent
autocorrelation functions, the turbulent spectrum, and the determination
of $\lambda_{\mathrm{AD}}$ are shown in Figures \ref{fig:iras16293_struct},
\ref{fig:iras16293_spectrum}, and \ref{fig:iras16293_ad}, respectively.
The dispersion and turbulent autocorrelation functions (Fig. \ref{fig:iras16293_struct})
display similar characteristics as those for Orion KL, but the differences
are made more obvious when considering the turbulence power spectrum.
Indeed, Figure \ref{fig:iras16293_spectrum} shows that the Kolmogorov-like
spectrum for IRAS 16293 scales as $\sim k^{-\left(1.8\pm0.3\right)}$,
while an inspection of Figure \ref{fig:iras16293_ad} makes it clear
that the apparent cut-off in the spectrum at $k\simeq0.4$ arcsec$^{-1}$
(vertical dotted line) is likely due to beam filtering. We therefore
only report an upper limit of $\lambda_{\mathrm{AD}}\lesssim1.8$
mpc for IRAS 16293 (assumed to be at a distance of 150 pc) for the
high frequency spectral cut-off due to ambipolar diffusion.

\subsection{NGC 1333 IRAS 4A}

The results of our dispersion analysis as applied to the NGC 1333
IRAS 4A polarization map of \citet{Girart2006} are shown in Figures
\ref{fig:ngc1333_struct}, \ref{fig:ngc1333_spectrum}, and \ref{fig:ngc1333_ad}.
For this source the Kolmogorov-like power spectrum is observed to
scale with $\sim k^{-\left(1.4\pm0.4\right)}$ (see Fig. \ref{fig:ngc1333_spectrum}),
while we find $k_{\mathrm{AD}}/2\pi\simeq0.66$ arcsec$^{-1}$ (or
$\lambda_{\mathrm{AD}}\simeq2.2$ mpc with an assumed distance of
300 pc) for the high frequency spectral cut-off due to turbulent ambipolar
diffusion (vertical dotted line in Fig. \ref{fig:ngc1333_ad}). However,
because of the weakness of $b^{2}\left(k\right)$ about $k_{\mathrm{AD}}$,
the latter's proximity to the spectral cut-off due to beam filtering,
and the aforementioned uncertainty in fitting the ordered component
of the turbulent autocorrelation function, we must acknowledge that
this estimate for $\lambda_{\mathrm{AD}}$ is tentative. As was the
case for Orion KL (and IRAS 16293 for the spectral index) these parameters
are consistent with expectations.

\section{Discussion\label{sec:Discussion}}

\subsection{Limitations of the Dispersion Technique}

The high spatial resolution with which the polarization data analyzed
in this paper were obtained has allowed us to determine fundamental
parameters that characterize the magnetized turbulence power spectrum
in some well-known star-forming regions. However, the same observing
mode that allows these realizations also brings with it some limitations
due to the filtering of low spatial frequencies inherent to interferometry.
We already discussed in detail in Section \ref{sec:Results} how this
impedes the precise determination of the power spectrum at low frequencies.
We now discuss two more consequences that result from this limitation.

\subsubsection{The Chandrasekhar-Fermi Technique}

As was shown in Paper I, the value for the turbulent to ordered magnetic
energy ratio $b^{2}\left(\ell=0\right)$ obtained when fitting the
dispersion function (see the broken curves in the top graphs of Figs.
\ref{fig:omc-1_struct}, \ref{fig:iras16293_struct}, and \ref{fig:ngc1333_struct})
corresponds to the value that one would normally use for the determination
of the ordered magnetic field strength with the Chandrasekhar-Fermi
equation \citep{CF1953}. More precisely, we have (see Eqs. {[}7{]}
and {[}8{]} of Paper I)

\begin{equation}
\left\langle B_{0}^{2}\right\rangle \simeq8\pi\rho\left(\frac{\sigma^{2}\left(v\right)}{b^{2}\left(0\right)}\right),\label{eq:CF}\end{equation}
where $\rho$ and $\sigma\left(v\right)$ are the mass density and
the one-dimensional turbulent velocity dispersion, respectively. However,
as was noted earlier the value of $b^{2}\left(\ell=0\right)$ determined
from interferometry data alone cannot be precisely determined because
of the filtering of low spatial frequencies. An error in the estimate
of the ordered magnetic field strength will then follow due to the
presence of $b^{2}\left(\ell=0\right)$ in the denominator of Equation
(\ref{eq:CF}). Although such an error will also arise with single-dish
data (where the filtering happens instead at high frequencies), the
relative importance of this error in comparison to what is expected
with interferometry observations can be studied by considering the
Fourier transform that links $b^{2}\left(\boldsymbol{\ell}\right)$
to $b^{2}\left(\mathbf{k}_{v}\right)$

\begin{equation}
b^{2}\left(\boldsymbol{\ell}\right)=\frac{1}{\left(2\pi\right)^{2}}\iint b^{2}\left(\mathbf{k}_{v}\right)e^{\boldsymbol{\ell}\cdot\mathbf{k}_{v}}d^{2}k_{v}.\label{eq:FT}\end{equation}

It follows from this and Equation (\ref{eq:b2(kv)}) that

\begin{eqnarray}
b^{2}\left(\ell=0\right) & = & \frac{1}{\left(2\pi\right)^{2}}\iint b^{2}\left(\mathbf{k}_{v}\right)d^{2}k_{v}\nonumber \\
 & = & \frac{1}{\left(2\pi\right)^{2}\left\langle \overline{B}^{2}\right\rangle }\iint\left\Vert H\left(\mathbf{k}_{v}\right)\right\Vert ^{2}\mathcal{R}_{\mathrm{t}}\left(\mathbf{k}_{v}\right)d^{2}k_{v},\label{eq:b2_0}\end{eqnarray}

\noindent which is valid in general. For a given magnetized turbulence
power spectrum $\mathcal{R}_{\mathrm{t}}\left(\mathbf{k}_{v}\right)$
the difference in $b^{2}\left(\ell=0\right)$ obtained with interferometry
and single-dish observations resides in the nature of the filtering
$H\left(\mathbf{k}_{v}\right)$ applied to the data. Although a determination
of $b^{2}\left(\ell=0\right)$ with single-dish will also be imprecise
because of the spectral filtering at higher frequencies, the error
is potentially more significant with interferometry since the spectral
filtering is concentrated at low frequencies (see the {}``visibility''
curves for Figs. \ref{fig:omc-1_spectrum}, \ref{fig:iras16293_spectrum},
and \ref{fig:ngc1333_spectrum}) where the turbulent spectrum $\mathcal{R}_{\mathrm{t}}\left(\mathbf{k}_{v}\right)$
peaks. Evidently, the relative importance in these errors will depend
on the precise shape of the corresponding single-dish and interferometer
dirty beams, as well as that of the underlying spectrum. Moreover,
the nature of the imprecision in the evaluation of $b^{2}\left(\ell=0\right)$
is made more complicated by the fact that $\left\langle \overline{B}^{2}\right\rangle \equiv\left\langle \overline{\mathbf{B}}\mathbf{\cdot}\overline{\mathbf{B}}\left(0\right)\right\rangle $,
present in the denominator of Equation (\ref{eq:b2_0}), will also
contain the same filtering integral as well as a similar one for the
ordered component of the magnetic field. In some cases this may alleviate
the aforementioned error, in others it may worsen it.

In the analysis of the SHARP OMC-1 data presented in Paper II the
effects of the beam filtering and signal integration through the thickness
of the cloud were corrected for by modeling the dispersion function
while assuming circular Gaussian magnetized turbulent autocorrelation
and beam functions. It is important to note, however, that even if
it were possible to completely remove the filtering due to $\left\Vert H\left(\mathbf{k}_{v}\right)\right\Vert ^{2}$
in Equation (\ref{eq:b2_0}), the corresponding value obtained for
$b^{2}\left(\ell=0\right)$ would still be underestimated because
of the aforementioned signal integration along the line of sight.
A correction for this effect would require a determination of the
turbulent correlation length (see Paper II), which can be obtained
by measuring the spectral width of $\mathcal{R}_{\mathrm{t}}\left(\mathbf{k}_{v}\right)$.
More precisely, the turbulent correlation length is inversely proportional
to the width of $\mathcal{R}_{\mathrm{t}}\left(\mathbf{k}_{v}\right)$.
As this could only be achieved in general if $\mathcal{R}_{\mathrm{t}}\left(\mathbf{k}_{v}\right)$
is known at low frequencies (where it peaks), it is apparent that
one would greatly benefit from combining single-dish and interferometry
observations to maximize the spectral coverage at both ends of the
spectrum.

\subsubsection{The Magnetized Turbulence Power Spectrum}

An inspection of the magnetized turbulence power spectrum $\mathcal{R}_{\mathrm{t}}\left(k\right)/\left\langle \overline{B}^{2}\right\rangle $
(or $\mathcal{R}_{1\mathrm{D}}\left(k\right)/\left\langle \overline{B}^{2}\right\rangle $)
shown in Figures \ref{fig:omc-1_spectrum}, \ref{fig:iras16293_struct},
and \ref{fig:ngc1333_spectrum} makes it clear that we are able to
determine its shape only at the high frequency end, whereas it is
expected that the inertial range of the power spectrum should extend
over several decades in length scale (or spatial frequency). Taking
the case of Orion KL as an example (see Fig. \ref{fig:omc-1_spectrum}),
and acknowledging the fact that, as discussed in Section \ref{sec:Results},
our spectrum is unreliably estimated on the low frequency end (i.e.,
for $k/2\pi\lesssim0.05$ arcsec$^{-1}$) we find that our analysis
uncovers much less than a decade of the underlying spectrum (i.e.,
$0.07\,\mathrm{arcsec}^{-1}\lesssim k/2\pi\lesssim0.2\,\mathrm{arcsec}^{-1}$).
One must therefore be cautious in putting too much weight in our determination
of the scaling laws characterizing the the small portion of the inertial
range probed with our observations of Orion KL, IRAS 16293, and NGC
1333 IRAS 4A. Correspondingly, we once again emphasize the benefits
that would thus be gained by combining single-dish and interferometry
data. That is, a single-dish map of suitable spatial extent would
ensure a good low frequency coverage, while a high-resolution interferometry
map of the same region would extend the measured turbulence power
spectrum far enough to precisely characterize the inertial range as
well as the turbulent ambipolar diffusion scale.

We should also keep in mind that the three-dimensional turbulence
power spectrum underlying our data in Equation (\ref{eq:autob2}),
i.e., $R_{\mathrm{3\mathrm{D},t}}\left(v,u\right)=\left\langle F\left(\mathbf{a},z\right)F\left(\mathbf{a}^{\prime},z^{\prime}\right)\right\rangle \left\langle \mathbf{B}_{\mathrm{t}}\left(\mathbf{a},z\right)\mathbf{B}_{\mathrm{t}}\left(\mathbf{a}^{\prime},z^{\prime}\right)\right\rangle $,
contains the autocorrelation of the (ordered) polarized emission as
well as that of the turbulent magnetic field. It therefore follows
that the magnetized turbulence power spectrum we extract from our
data does not exactly correspond to that of the turbulent magnetic
field, but is somewhat broadened by the polarized emission spectrum.
The importance of this effect may be advantageously investigated through
numerical analyses and simulations.

\subsection{Further Improvements and Applications\label{sub:future}}

In all of our applications of the dispersion technique (i.e., in Paper
I, II, and here) we always treated turbulence as being isotropic with
a Kolmogorov-like power spectrum. This is, of course, a simplification
that we do not expect to hold for magnetized turbulence in a weakly
ionized plasma due to the anisotropy of motions in directions parallel
and perpendicular to the magnetic field brought about by the Lorentz
force. If for example one considers the theory of \citet{Goldreich1995}
for incompressible magnetized turbulence, then different power law
scaling are expected for the power spectra measured along these two
distinct orientations. Such anisotropy has, in fact, been measured
in Taurus by \citet{Heyer2008} through $^{12}$CO $\left(J=1\rightarrow0\right)$
observations and optical polarization measurements using principal
component analysis.

It would be straightforward in principle to extend our dispersion
technique to allow for the detection of anisotropy. We would simply
have to \emph{locally} determine the mean orientation of the magnetic
field about the position of a given datum on a polarization map and
define, say, two sets of displacements $\ell_{\perp}$ and $\ell_{\parallel}$
depending whether the distance vector $\boldsymbol{\ell}$ linking
that point to another one on the map is oriented more or less perpendicular
or parallel to the mean magnetic field, respectively. Two dispersion
analyses could then be performed, one for each of the $\ell_{\perp}$
and $\ell_{\parallel}$ data sets. The main constraint in applying
this technique for the polarization maps analyzed in this paper is
the lack of data points. One needs a large number of points in order
to accurately estimate the dispersion function. Moreover, it is also
imperative that the change in orientation of the polarization vectors
on a map varies smoothly enough that a mean direction for the magnetic
field can be adequately calculated. Accordingly, we plan to attempt
the implementation of this technique on the SHARP OMC-1 polarization
map presented in Paper II in a future publication.

Another natural extension of this technique concerns the analysis
of polarization maps of face-on spiral galaxies (see, e.g., the map
of M51 presented in \citet{Fletcher2010}). Although the polarization
measured for external galaxies is not due to emission from anisotropic
grains but from synchrotron radiation, we see no reason why the dispersion
technique could not be applied to such cases. It would then also be
natural to study the aforementioned anisotropy of magnetized turbulence
since face-on spiral galaxies often show a mean magnetic field orientation
that closely traces the spiral arms \citep{Fletcher2010}. Such analyses
would also provide a detailed study of magnetized turbulence on a
much larger scale than we have achieved so far (i.e., on galactic
scales instead of that of molecular clouds).

\section{Summary\label{sec:Summary}}

We presented an application of our magnetic field dispersion technique
to high spatial resolution SMA polarization maps obtained for Orion
KL in OMC-1, IRAS 16293, and NGC 1333 IRAS 4A. We showed how one can
take advantage of such high resolution data to characterize the magnetized
turbulence power spectrum in the inertial and dissipation ranges.
For Orion KL we determine that the inertial range of the spectrum
approximately scales with $k^{-\left(2.9\pm0.9\right)}$ and we report
a value of 9.9 mpc for $\lambda_{\mathrm{AD}}$, the high spatial
frequency cutoff presumably due to turbulent ambipolar diffusion.
For the same parameters we have $\sim k^{-\left(1.4\pm0.4\right)}$
and a tentative value of $\lambda_{\mathrm{AD}}\simeq2.2$ mpc for
NGC 1333 IRAS 4A, and $\sim k^{-\left(1.8\pm0.3\right)}$ and an upper
limit $\lambda_{\mathrm{AD}}\lesssim1.8$ mpc for IRAS 16293.

\acknowledgements{The authors thank the referee, P. M. Koch, for his valuable comments,
which greatly improved this paper. M.H.'s research is funded through
the NSERC Discovery Grant, Canada Research Chair, Canada Foundation
for Innovation, Ontario Innovation Trust, and Western's Academic Development
Fund programs. }

\appendix
\section{Wiener Filter and Data Processing}\label{sec:Wiener}

If we take into account the contribution of the spectral noise $n\left(\mathbf{k}_{v}\right)$
to $b^{2}\left(\mathbf{k}_{v}\right)$ we can write

\begin{equation}
b^{2}\left(\mathbf{k}_{v}\right)=\hat{b}^{2}\left(\mathbf{k}_{v}\right)+n\left(\mathbf{k}_{v}\right),\label{eq:b2+n}\end{equation}

\noindent where the function

\begin{equation}
\hat{b}^{2}\left(\mathbf{k}_{v}\right)=\left\Vert H\left(\mathbf{k}_{v}\right)\right\Vert ^{2}\frac{\mathcal{R}_{\mathrm{t}}\left(\mathbf{k}_{v}\right)}{\left\langle \overline{B}^{2}\right\rangle }\label{eq:b^2}\end{equation}
is assumed noiseless. The well-known solution for the Wiener filter
to be applied to $b^{2}\left(\mathbf{k}_{v}\right)$ in order recover
$\mathcal{R}_{\mathrm{t}}\left(\mathbf{k}_{v}\right)/\left\langle \overline{B}^{2}\right\rangle $
(in the least-squared sense) is given by \citep{Press1992} 

\begin{equation}
\phi\left(\mathbf{k}_{v}\right)=\frac{\left\Vert \hat{b}^{2}\left(\mathbf{k}_{v}\right)\right\Vert ^{2}/\left\Vert H\left(\mathbf{k}_{v}\right)\right\Vert ^{2}}{\left\Vert \hat{b}^{2}\left(\mathbf{k}_{v}\right)\right\Vert ^{2}+\left\Vert n\left(\mathbf{k}_{v}\right)\right\Vert ^{2}}.\label{eq:phi_1}\end{equation}

It is then necessary to somehow estimate the noise level $n\left(\mathbf{k}_{v}\right)$
and insert Equation (\ref{eq:b^2}) into Equation (\ref{eq:phi_1})
to express the filter as a function of $\left\Vert H\left(\mathbf{k}_{v}\right)\right\Vert ^{2}$
and the signal-to-noise ratio, as is often done. It is straightforward
to show that the Wiener filter for our problem can be expressed as 

\begin{equation}
\phi\left(\mathbf{k}_{v}\right)=\frac{\left\Vert H\left(\mathbf{k}_{v}\right)\right\Vert ^{2}}{\left\Vert H\left(\mathbf{k}_{v}\right)\right\Vert ^{4}+\frac{\left\Vert n\left(\mathbf{k}_{v}\right)\right\Vert ^{2}}{\left\Vert \mathcal{R}_{\mathrm{t}}\left(\mathbf{k}_{v}\right)\right\Vert ^{2}/\left\langle \overline{B}^{2}\right\rangle ^{2}}}.\label{eq:phi_final}\end{equation}

\noindent Equation (\ref{eq:phi_final}) is seen to tend to the obvious
limit of $\phi\left(\mathbf{k}_{v}\right)=1/\left\Vert H\left(\mathbf{k}_{v}\right)\right\Vert ^{2}$
when $n\left(\mathbf{k}_{v}\right)$ vanishes. We compute our Wiener
filter by \emph{i)} determining the mean level $\overline{n}$ for
$n\left(\mathbf{k}_{v}\right)$ in the high frequency end of the spectrum
for $b^{2}\left(\mathbf{k}_{v}\right)$ where $\hat{b}^{2}\left(\mathbf{k}_{v}\right)$
is negligible, \emph{ii)} subtracting $\overline{n}$ from $b^{2}\left(\mathbf{k}_{v}\right)$
to approximately obtain $\mathcal{R}_{\mathrm{t}}\left(\mathbf{k}_{v}\right)/\left\langle \overline{B}^{2}\right\rangle $,
and \emph{iii)} inserting $\left\Vert \overline{n}\right\Vert ^{2}$
(in lieu of $\left\Vert n\left(\mathbf{k}_{v}\right)\right\Vert ^{2}$)
and $\left\Vert \mathcal{R}_{\mathrm{t}}\left(\mathbf{k}_{v}\right)\right\Vert ^{2}/\left\langle \overline{B}^{2}\right\rangle ^{2}$
in Equation (\ref{eq:phi_final}). 

\begin{figure}
\epsscale{1.0}\plotone{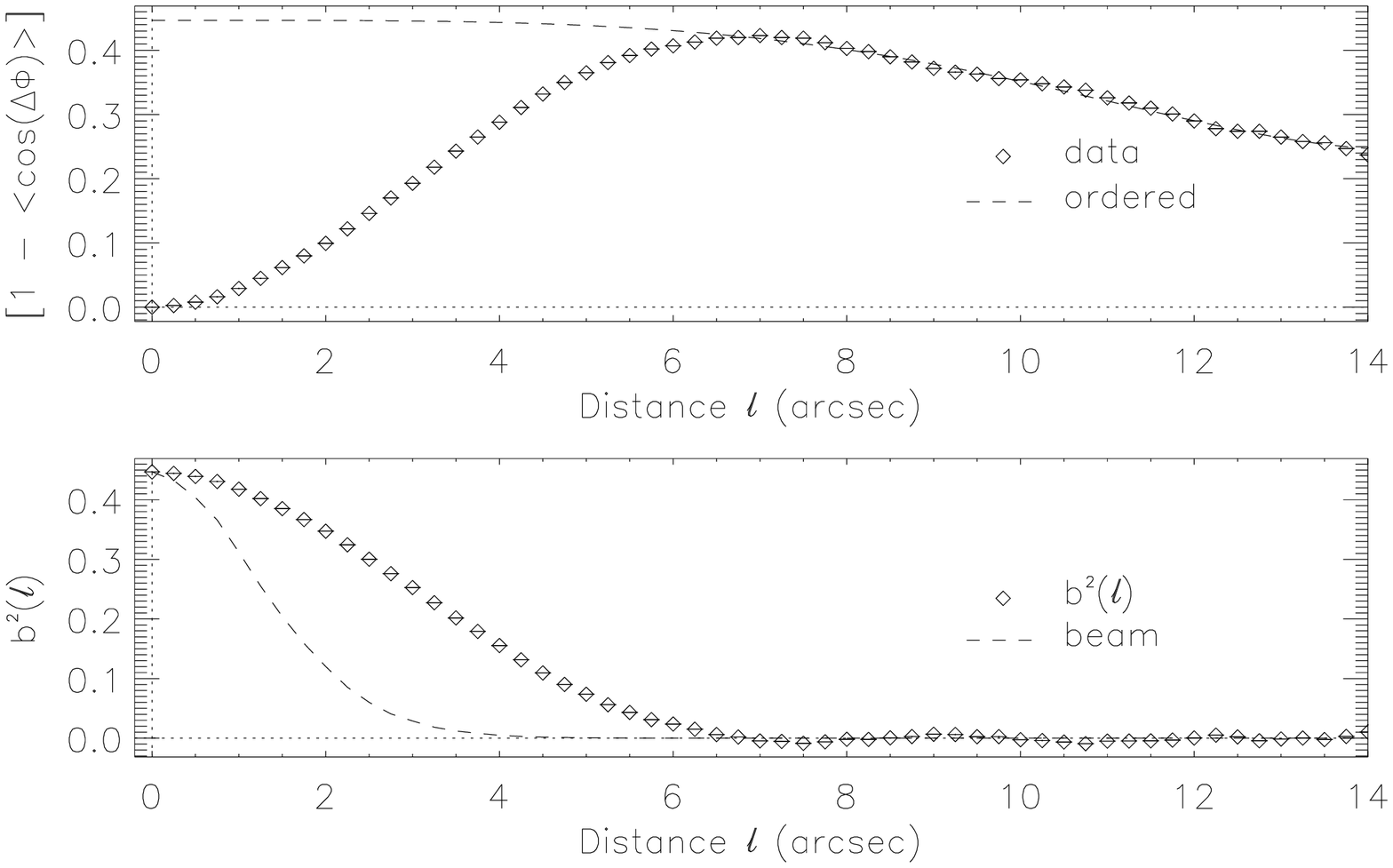}

\caption{\label{fig:omc-1_struct}\emph{Top:} The dispersion function $1-\left\langle \cos\left[\Delta\Phi\left(\ell\right)\right]\right\rangle $
for Orion KL. The broken curve ({}``ordered'') is the fit for the
sum of the turbulent to total magnetic energy ratio $b^{2}\left(0\right)$
and the ordered component $\sum_{j=1}^{3}a_{2j}\ell^{2j}$ implicit
to the data (symbols) using the points where $\ell\geq7\arcsec$.
Both functions are plotted as a function of $\ell$. \emph{Bottom:}
the turbulent autocorrelation function $b^{2}\left(\ell\right)$ (symbols),
as obtained by subtracting the data points to the {}``ordered''
curve in the top graph, while the broken curve shows the radial profile
of the {}``mean autocorrelated synthesized beam'' (see text). }

\end{figure}

\begin{figure}
\plotone{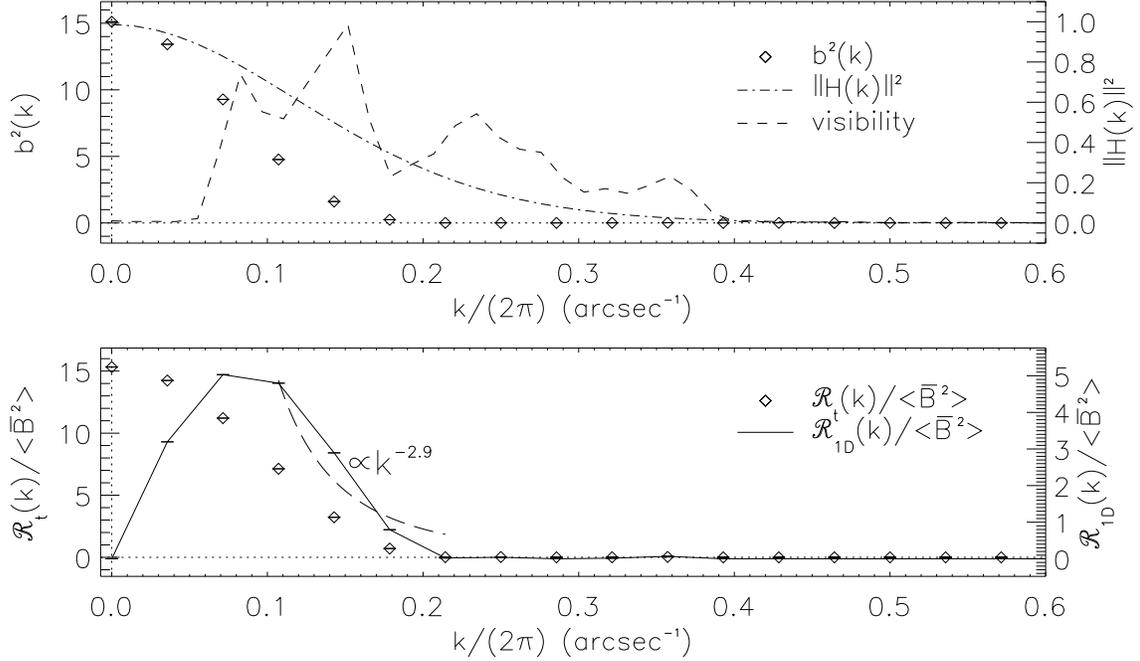}

\caption{\label{fig:omc-1_spectrum}\emph{Top:} spectra associated to $b^{2}\left(\ell\right)$
(i.e., $b^{2}\left(k\right)$ with $k=\left|\mathbf{k}_{v}\right|$;
symbols) and the mean autocorrelated synthesized beam (i.e., $\left\Vert H\left(k\right)\right\Vert ^{2}$)
calculated by taking the Fourier transform of the corresponding functions
shown in the bottom graph of Figure \ref{fig:omc-1_struct}. The visibility
data (or the spectrum profile of the dirty beam) is also shown (see
text) and is normalized to its peak level (using the scale on the
right). \emph{Bottom:} our results for the magnetized turbulence power
spectrum profile $\mathcal{R}_{\mathrm{t}}\left(k\right)/\left\langle \overline{B}^{2}\right\rangle $
(symbols) and the associated one-dimensional {}``Kolmogorov-like''
spectrum $\mathcal{R}_{1\mathrm{D}}\left(k\right)/\left\langle \overline{B}^{2}\right\rangle $.
The broken curve shows an approximate power law fit $k^{-\left(2.9\pm0.9\right)}$
to the inertial range.}

\end{figure}

\begin{figure}
\plotone{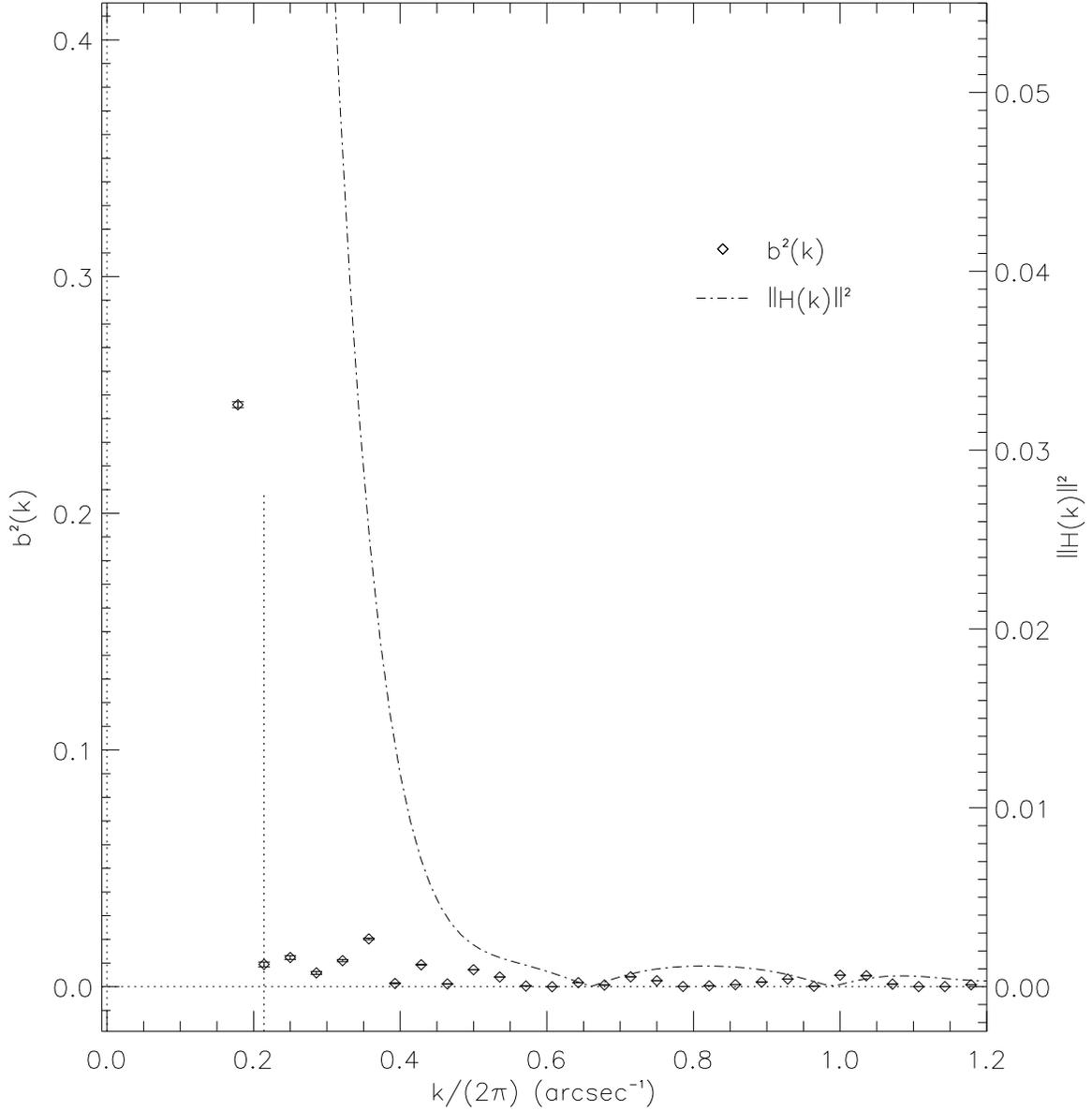}

\caption{\label{fig:omc-1_ad}Magnification of the vertical scale of the top
graph of Figure \ref{fig:omc-1_spectrum}. For Orion KL we detect
$k_{\mathrm{AD}}/2\pi\simeq0.22$ arcsec$^{-1}$ (or $\lambda_{\mathrm{AD}}\simeq9.9$
mpc with an assumed distance of 450 pc) for the high frequency spectral
cut-off, which is probably due to turbulent ambipolar diffusion (vertical
dotted line).}

\end{figure}

\begin{figure}
\plotone{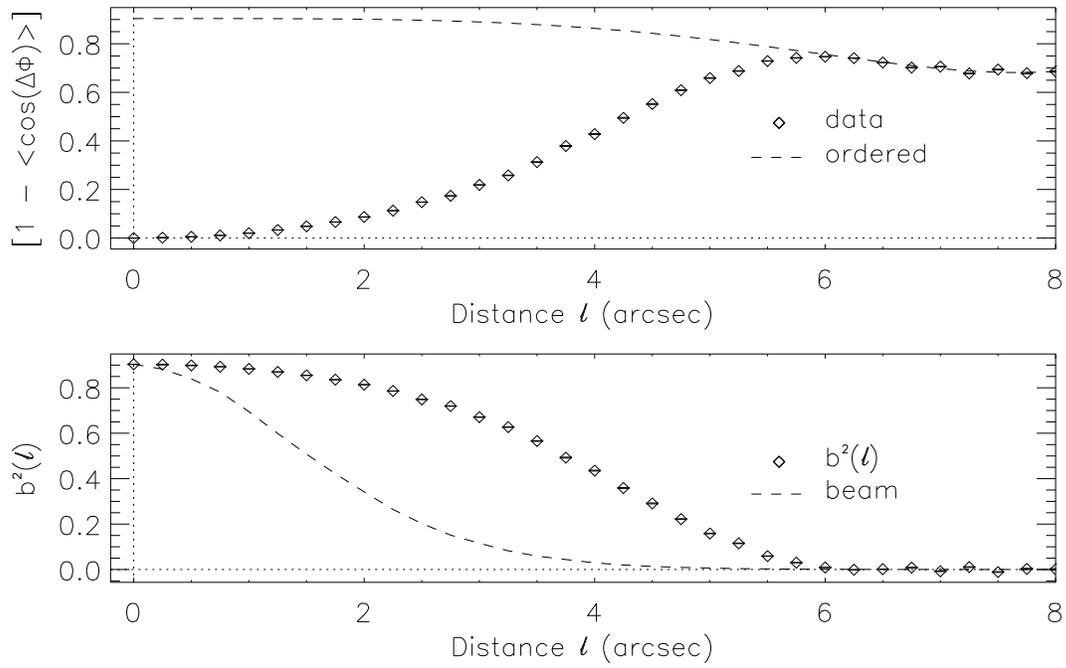}

\caption{\label{fig:iras16293_struct}Same as Figure \ref{fig:omc-1_struct}
but for IRAS 16293; data points where $\ell\geq6.25\arcsec$ were
used for the fit on the top graph (broken curve).}

\end{figure}

\begin{figure}
\plotone{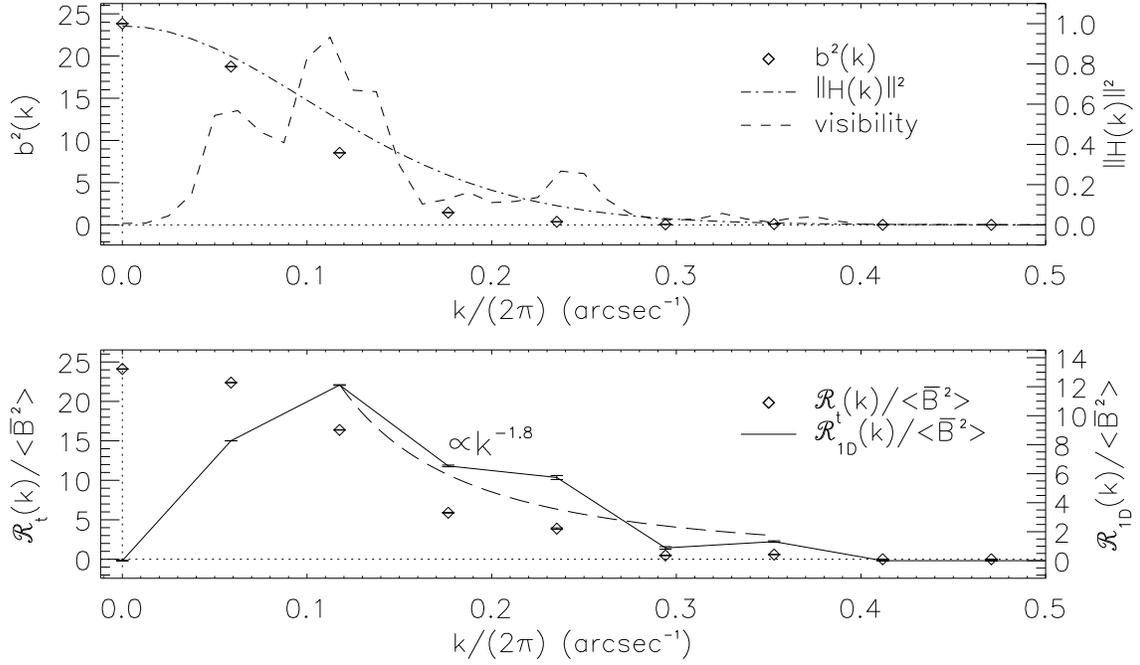}

\caption{\label{fig:iras16293_spectrum}Same as Figure \ref{fig:omc-1_spectrum}
but for IRAS 16293. However, for this source the broken curve on the
bottom graph shows an approximate power law fit $k^{-\left(1.8\pm0.3\right)}$
to the inertial range.}

\end{figure}

\begin{figure}
\plotone{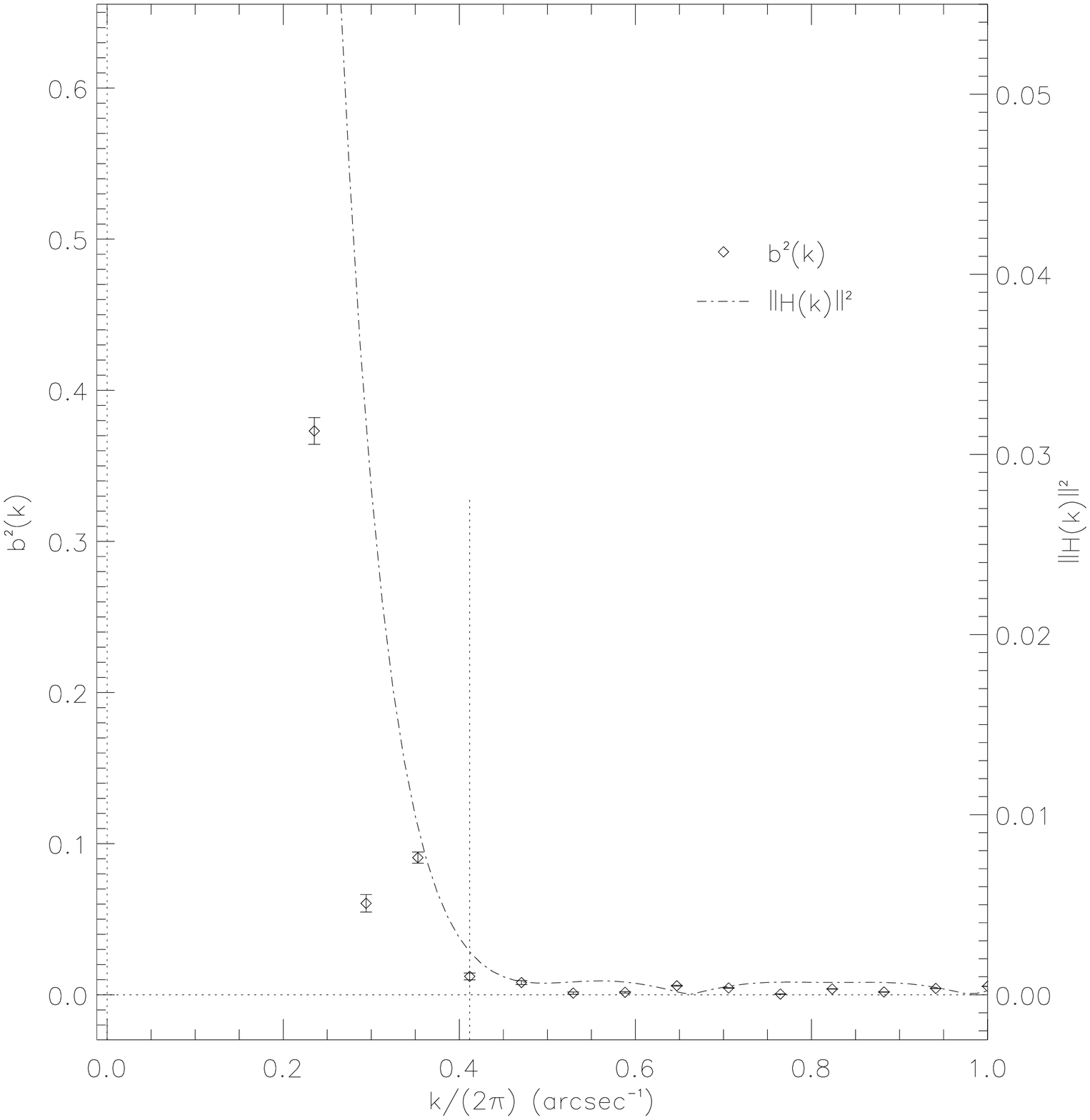}

\caption{\label{fig:iras16293_ad}Same as Figure \ref{fig:omc-1_ad} but for
IRAS 16293. However, for this source we observe that the high frequency
cut off present in the data for $b^{2}\left(k\right)$ (i.e., before
applying the Wiener filter) is correlated with the bandwidth subtended
by the synthesized beam$\left\Vert H\left(k\right)\right\Vert ^{2}$,
which cuts off at $k/2\pi\simeq0.41$ arcsec$^{-1}$. We therefore
have an upper limit of $\lambda_{\mathrm{AD}}\lesssim1.8$ mpc for
IRAS 16293 (assumed distance of 150 pc) for the high frequency spectral
cut-off expected from turbulent ambipolar diffusion (vertical dotted
line).}

\end{figure}

\begin{figure}
\plotone{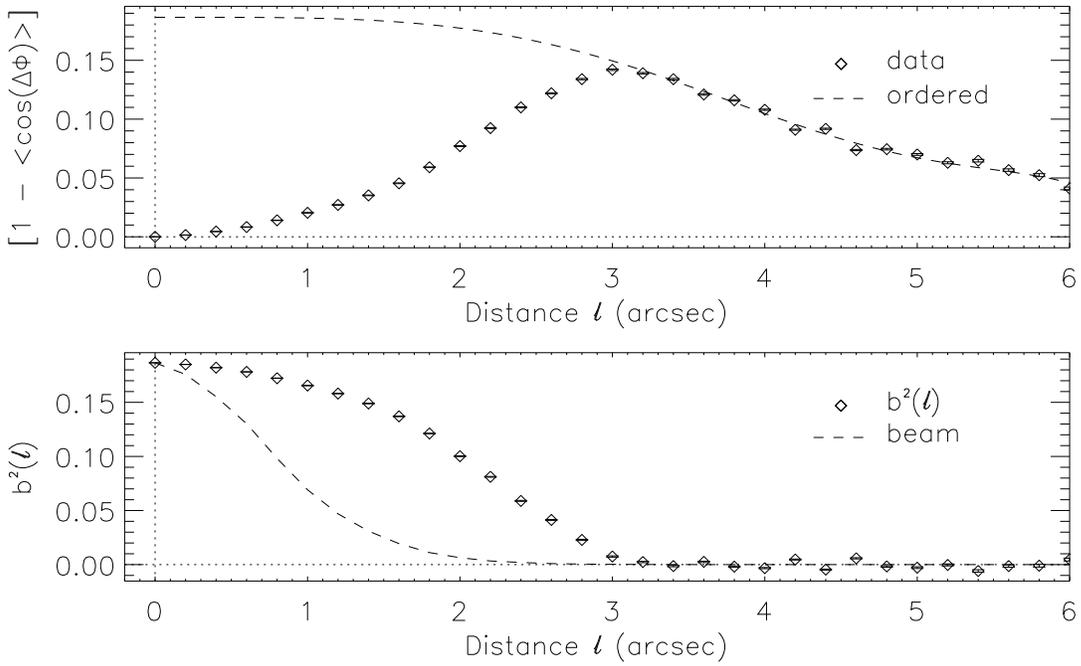}

\caption{\label{fig:ngc1333_struct}Same as Figure \ref{fig:omc-1_struct}
but for NGC 1333 IRAS 4A. The broken curve on the top graph ({}``ordered'')
is the fit for the turbulent to total magnetic energy ratio $b^{2}\left(0\right)$
and the ordered component $\sum_{j=1}^{4}a_{2j}\ell^{2j}$ implicit
to the data (symbols) using the points where $\ell\geq3\farcs4$. }

\end{figure}

\begin{figure}
\plotone{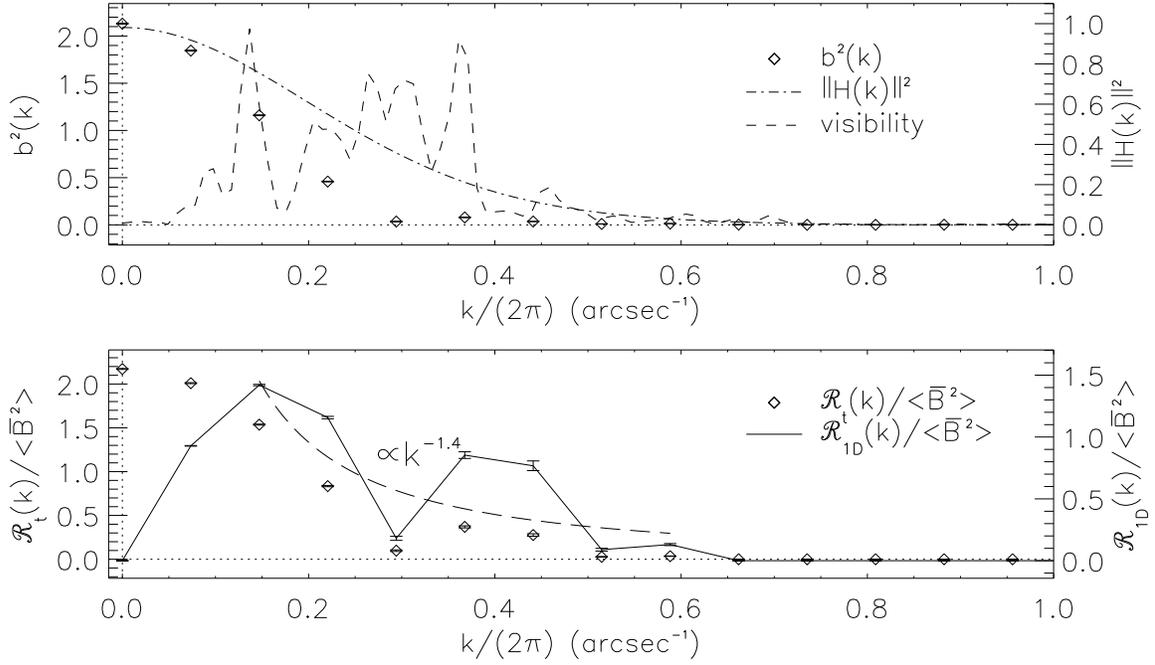}

\caption{\label{fig:ngc1333_spectrum}Same as Figure \ref{fig:omc-1_spectrum}
but for NGC 1333 IRAS 4A. However, for this source the broken curve
on the bottom graph shows an approximate power law fit $k^{-\left(1.4\pm0.4\right)}$
to the inertial range.}

\end{figure}

\begin{figure}
\plotone{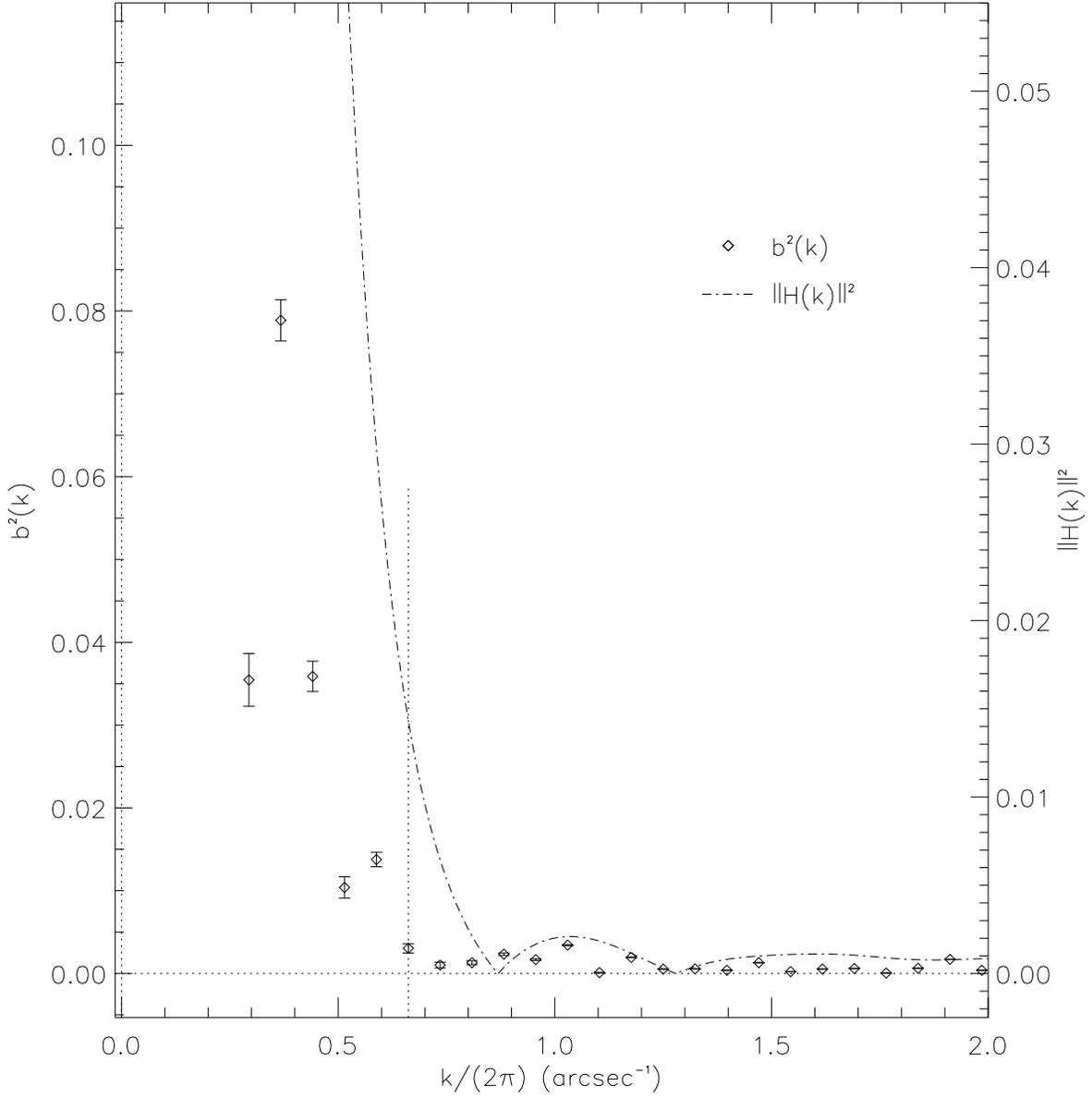}

\caption{\label{fig:ngc1333_ad}Same as Figure \ref{fig:omc-1_ad} but for
NGC 1333 IRAS 4A. However, for this source we tentatively detect $k_{\mathrm{AD}}/2\pi\simeq0.66$
arcsec$^{-1}$ (or $\lambda_{\mathrm{AD}}\simeq2.2$ mpc with an assumed
distance of 300 pc) for the high frequency spectral cut-off, which
is probably due to turbulent ambipolar diffusion (vertical dotted
line).}

\end{figure}

Finally, we note that we processed the data in the spectral domain
using discrete Fourier transforms (DFT). Although the shapes of the
$b^{2}\left(\ell\right)$ functions shown in Figures \ref{fig:omc-1_struct},
\ref{fig:iras16293_struct}, and \ref{fig:ngc1333_struct} are such
that they tend to smoothly approach zero at the larger values of $\ell$,
the presence of noise and of residual levels can potentially be the
cause of {}``edge effects'' and ensuing contamination of the spectra
$b^{2}\left(\mathbf{k}_{v}\right)$. We have therefore windowed the
data and the synthesized beam in the $\ell$ domain with a Hanning
window \citep{Hamming1997} before applying DFTs in order to minimize
these effects.

\end{document}